\documentclass[useAMS,usenatbib,fleqn]{mnras}
\usepackage{graphicx}	
\usepackage{amsmath}	
\usepackage{amssymb}	
\usepackage{multicol}        
\usepackage{bm}		
\usepackage[T1]{fontenc}
\usepackage{ae,aecompl}
\usepackage{times}
\usepackage{color}
\def\be{\begin{equation}}
\def\ee{\end{equation}}
\def\ba#1\ea{\begin{align}#1\end{align}}
\newcommand{\vs}{\nonumber\\}
\newcommand{\rhob}{\bar\rho}
\newcommand{\refeq}[1]{Eq.~(\ref{eq:#1})}          
\newcommand{\refeqs}[2]{Eqs.~(\ref{eq:#1})--(\ref{eq:#2})}
\def\d{\delta}
\def\vk{\vec{k}}
\def\vx{\vec{x}}
\def\vq{\vec{q}}
\def\vs{\vec{s}}
\def\abg{a_\mathrm{bg}}
\def\dotabg{\dot{a}_\mathrm{bg}}
\def\ddotabg{\ddot{a}_\mathrm{bg}}
\def\Hbg{H_\mathrm{bg}}
\def\Hbgn{H_\mathrm{bg,0}}
\def\rhobg{\rho_\mathrm{bg}}

\def\comment#1{}

\newcommand{\MPA}{Max Planck Institute for Astrophysics}

\newcommand{\Gadget}{\textsc{gadget}}
\newcommand{\gadget}{\textsc{gadget}4\,}

\newcommand{\as}[1]{\textcolor{black}{#1}}

\title[Large Scale Tidal Field]{Cosmological N-Body Simulations with a Large-Scale Tidal Field}
\author[A. S. Schmidt et al.]{Andreas. S. Schmidt$^{1}$\thanks{E-mail: aschmidt@mpa-garching.mpg.de} , Simon D. M. White$^{1}$, Fabian Schmidt$^{1}$ and Jens St\"ucker$^{1}$
\\
$^{1}$\MPA , Karl-Schwarzschild-Str. 1, 85741 Garching, Germany\\}

\begin{document}

\date{Accepted . Received ; in original form 2018 }

\pagerange{\pageref{firstpage}--\pageref{lastpage}} \pubyear{2018}

\maketitle

\label{firstpage}

\begin{abstract}
In this paper we carry out anisotropic ``separate universe'' simulations by including a large-scale tidal field in the N-body code \gadget using an anisotropic expansion factor $A_{ij}$.
We use the code in a pure \textit{particle-mesh} (PM) mode to simulate the evolution of 16 realizations of an initial density field with and without a large-scale tidal field, which are then used to measure the \textit{response function} describing how the tidal field influences structure formation in the linear and non-linear regimes. Together with the previously measured response to a large scale overdensity, this completely describes the nonlinear matter bispectrum in the squeezed limit. 
We find that, contrary to the density response, the tidal response never significantly exceeds the large-scale perturbation-theory prediction even on nonlinear scales for the redshift range we discuss.
We develop a simple halo model that takes into account the effect of the tidal field and compare it with our direct measurement from the anisotropic N-body simulations.
\end{abstract}

\section{Introduction}
\label{sec:Intro}
Modern large-scale galaxy surveys offer a precise measurement of the density distribution of galaxies and matter, using a variety of probes like baryon acoustic oscillations (BAO), redshift space distortions (RSD), and gravitational lensing.
With this data they aim to understand the cause of the accelerated expansion, and the physics of the early universe (e.g. Inflation), as well as to measure the curvature of the universe and the nature of primordial fluctuations.

This information is normally inferred from n-point statistics that compress the information contained in the underlying field.
The simplest of these statistics is the two-point correlation function $\xi(\vec{x})$ with its Fourier counterpart, the power spectrum $P(k) \propto \left\langle \delta(\vec{k}) \delta(\vec{k}')\right\rangle$. Given the initial conditions provided by the cosmic microwave background (CMB), this provides a possibility to constrain the time-evolution of structure in the universe.
At early times, when linear perturbation theory accurately describes the structure evolution of large scales, the power spectrum does fully specify the underlying field.
However at late times, when structure formation becomes non-linear, at least in the standard $\Lambda$CDM model, perturbation theory breaks down and cannot fully describe the structure seen in galaxy surveys.

To unleash the full potential of large-scale galaxy surveys, a better understanding of the non-linear evolution is necessary.
In finite volume surveys there are effects from large-scale perturbations which are not directly observable.
These fluctuations, even though they have small amplitudes, modify structure on smaller scales due to the non-linear mode coupling that needs to be included in the analysis.
There are two leading effects that come into play.
The first is due to a coherent large-scale over- or underdensity in which the survey volume is embedded.
The effect of a change in overdensity has been well studied using ``separate universe simulations,'' N-body simulations with a modified set of cosmological parameters implementing the gravitational effect of the large-scale overdensity \cite[e.g.][]{Frenk/White/Davies:1988,mcdonald:2003,sirko:2005,martino/sheth:2009,gnedin/kravtsov/rudd:2011,li/hu/takada:2014,wagner/etal:2014}. The second effect is a large-scale tidal field, which will make the local statistics anisotropic. 
The effects from such a field have not been studied in the quasi-linear and nonlinear regime, while for the linear regime an expression for the influence has been obtained from second-order perturbation theory \citep{2017PhRvD..95h3522A,2017JCAP...06..053B} and has been further studied in this regime by \cite{2017arXiv171100018L}, and \cite{2017arXiv171100012A}.
The change in the angle dependent 3D power spectrum is quantified by a \textit{response function} which is independent of wavenumber in the linear regime.

\cite{2017PhRvD..95h3522A} showed that large-scale tides produce an anisotropic redshift space power spectrum which mimics RSD from peculiar velocities and the Alcock-Paczynski distortion.
To linear order the large-scale tidal field, being a quadrupole, does not impact the angle averaged one-dimensional power spectrum and only weakly affects the angle-averaged redshift space power spectrum.
To measure the effects of tidal field directly, the angle dependent three-dimensional power spectrum $P(\vec{k})$ has to be used.
\begin{figure*}
\includegraphics[width=0.95\textwidth]{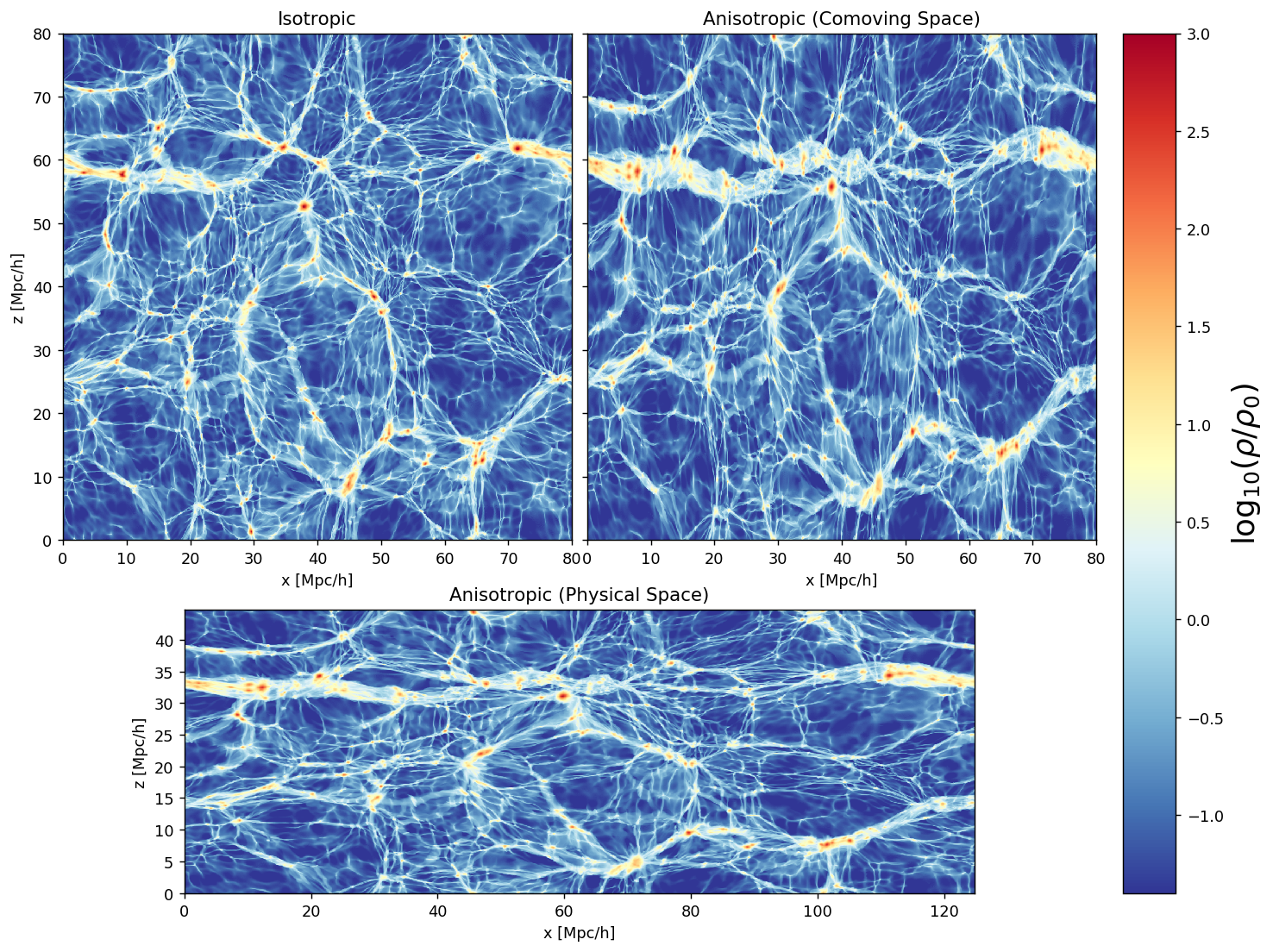}
\caption{Thin slices of the density field of a sample simulation with a boxsize of 80 cMpc/h and a strong tidal field with $\lambda = (-0.5, 0, 0.5)$. The left upper panel shows a standard simulation without large-scale tidal field.
The right upper panel shows the same initial conditions evolved using a strong tidal field in comoving space.
Here we see that most halos seem more elliptical than spherical and some structures are merged that are still separated in the standard case along certain axes.
The lower panel shows the tidal field simulation in physical space, where the axes are rescaled according to the anisotropic scale factors.
In physical space most halos appear spherical but on larger scales there is a clear alignment of structure with the tidal field.
The color represents the overdensity as given by the colorbar on the right.}
\label{fig:image_density_field}
\end{figure*}
The anisotropy of the power spectrum also contains information on super-horizon perturbations \citep{Byrnes/Domenech/Sasaki:2016} or statistical anisotropies in the two-point correlation function originating from physics of inflation \citep{Jeong/Kamionkowski:2012}. To disentangle these primordial effects from the late-time effects of tidal fields requires an accurate understanding of the latter. 
Further, the tidal response is an important ingredient in the covariance of the nonlinear matter power spectrum \citep{bertolini/etal,bertolini/solon,mohammed/etal,2017JCAP...06..053B,barreira/schmidt:2}. Finally, \citet{Barreira/Krause/Schmidt:2017} recently derived the super-sample covariance of weak lensing power spectra using the response function approach.
They showed that the super-sample covariance contains significant contributions from the tidal field response.
Hence, these have to be included in any cosmic shear analysis.
To address these issues, we have therefore modified the particle-mesh (PM) part of the code \gadget, to allow us to measure the response to large-scale tidal fields up to $k \sim 2$~h/cMpc, where here and throughout cMpc stands for Mpc in comoving coordinates, and hence into the non-linear regime. 
In the linear regime, we find that the measured response follows the theoretical prediction, but in the mildly non-linear regime, we find a substantial suppression of the response with respect to this linear prediction.

We structure our paper as follows.
In Sec. \ref{sec:model} we set out our description of the tidal field and its equivalence to an anisotropically expanding universe.
Sec. \ref{sec:Nbody} describes the implementation into \gadget. 
We then discuss the definition of the response function we are going to measure and compare to theory in Sec. \ref{sec:theo}.
Further in Sec. \ref{sec:halomodel} we consider predictions for the response function, which includes that of a simple halo model.
The simulation setup is described in Sec. \ref{sec:sims}, which is followed by a description of how we measure the response in the simulations.
Sec. \ref{sec:results} contains our results and some discussion.
Finally, in Sec. \ref{sec:summary} we summarize our findings.

\section{The Large-Scale Tidal Field}
\label{sec:model}
The following describes the equations used to simulate a portion of the universe which is embedded in a large-scale tidal field field. 

Consider an FRW metric perturbed by a long-wavelength potential perturbation $\Phi$, which is defined as the perturbation to the 00 component of the metric. Its leading locally observable effects are described by the corresponding tidal tensor
\begin{align}
  \partial_i\partial_j\Phi(\vx,t) &= 4\pi G\rhobg\,\Pi_{ij} \\
  \Pi_{ij} &= K_{ij,L} + \frac13 \delta_L \delta_{ij} = \frac{\partial_{i}\partial_{j}}{\nabla^2} \delta_L\,,
  \label{eq:Pi}
\end{align}
where $\delta_L, K_{ij,L}$ are the long-wavelength density and tidal perturbation corresponding to $\Phi$, respectively. Now consider the case where the wavelength of this mode is much larger than the size of the simulation box. Then, we can approximate $\Pi_{ij}$ as spatially (but not temporally) constant. If $\Pi_{ij}\propto \d_{ij}$, equivalently $K_{ij,L}=0$, the long-wavelength density perturbation can be absorbed in modified cosmological parameters, \as{as derived in
\cite{baldauf/etal:2011,CFCpaper2}
    and applied to simulations in \cite{
      sirko:2005,gnedin/kravtsov/rudd:2011,li/hu/takada:2014,wagner/etal:2014}.}
That is, even in the presence of the long-wavelength perturbation $\d_L$, the background metric within the simulation retains its Friedmann-Robertson-Walker (FRW) form,
\ba
ds^2 = -dt^2 + a^2(\tau) (1 + {\rm K} r^2 / 4)^{-2}  \d_{ij} dx^i dx^j\,,
\label{eq:FRW}
\ea
where ${\rm K}$ is the curvature, and both $a(\tau)$ and ${\rm K}$ are modified by the long-wavelength density perturbation.

In this paper, we are interested in the case $K_{ij,L} \neq 0$. 
Consider a homogeneous but anisotropic expanding spacetime,
\ba
ds^2 = -dt^2 + A_{ij}(t) A^j_{\  k}(t) dx^i dx^k\,,
\label{eq:anisoFRW}
\ea
where we will also write 
\ba
\label{eq:anisoScaleDef}
A_{ij}(t) = \abg(t) \alpha_{ij}(t)\,,
\ea
where $\alpha_{ij}$ is a symmetric matrix encoding the scale factor perturbation and $\abg$ is an isotropic ``background'' scale factor which we will specify later. \refeq{anisoFRW} is formally the metric describing a Bianchi~I spacetime. As shown in \cite{ip/schmidt:2016} however, a Bianchi~I spacetime is \emph{not} equivalent to an FRW spacetime with a tidal perturbation. Indeed, in order to source the $\alpha_{ij}$ in \refeq{anisoScaleDef}, a significant anistropic stress is necessary, which is not present in standard N-body simulations containing only non-relativistic matter.

However, since motions in large-scale structure are non-relativistic, one can still use \refeq{anisoFRW} to \emph{simulate} the effect of a long-wavelength tidal field. The spatially homogeneous metric \refeq{anisoFRW} offers the advantage  of being compatible with the periodic boundary conditions employed in N-body simulations. For this, we choose $\alpha_{ij}(t)$ to match the time-time-component of the metric in the comoving (Fermi) frame of the particles induced by a long-wavelength tidal field $\Pi_{ij}(t)$. This approach is related to the ``fake separate universe'' approach considered by \cite{hu/etal:2016} and \cite{chiang/etal:2016} for isotropic isocurvature perturbations due to dark energy and/or neutrinos.

In order to derive this matching for a general time dependence of the long-wavelength tidal field, we consider the geodesic deviation. Particle trajectories can be written as
\be
\vx = \vq + \vs(\vq,t)\,,
\label{eq:vx}
\ee
where all coordinates are comoving with respect to $\abg$, $\vq$ is the initial position and $\vs(\vq,0)=0$. For non-relativistic particles in a perturbed FRW spacetime with scale factor $\abg$, the displacement obeys
\be
\ddot{\vs} + 2 \Hbg \dot{\vs} = - \boldsymbol{\nabla}_x \Phi(\vq+\vs)\,,
\ee
where $\Hbg = \dotabg/\abg$, and $\boldsymbol{\nabla}_x$ indicates the gradient with respect to the comoving coordinate \refeq{vx}. Taking the derivative of this equation with respect to $\vq$ yields the evolution of the geodesic deviation $M_{ij} \equiv \partial_{q,j} s_i$:
\be
\ddot{M}_{ij} + 2 \Hbg \dot{M}_{ij} = - \left( \d_j^{\  k} + M_j^{\  k}\right) \partial_{x,k}\partial_{x,i} \Phi\,.
\label{eq:Mdot}
\ee
Now consider the motion of comoving test particles in an unperturbed anisotropic spacetime \refeq{anisoFRW}. In terms of physical coordinates, their acceleration is
\be
\ddot{r}_i = \frac{d^2}{dt^2} (\abg \alpha_{ij} )\; x^j\,,
\ee
where $x^j$ is the comoving coordinate with respect to the metric \refeq{anisoFRW}, which is constant for comoving observers. On the other hand, in terms of a fictitious FRW spacetime described by $\abg(t)$, we have $r_i = \abg (q_i + s_i)$, so that this trajectory corresponds to a Lagrangian displacement of
\ba
\ddot{r}_i &= \frac{d^2}{dt^2} \left[\abg(q_i+s_i)\right] \nonumber\\
&= \ddotabg(q_i + s_i) + 2\dotabg \dot{s}_i + \abg \ddot{s}_i\,.
\ea
Equating the previous two equations, and using the relation $\alpha_{ij} x^j = q_i+s_i$, we obtain
\be
\ddot{s}_i + 2\Hbg \dot{s}_i = \left[2\Hbg \dot{\alpha}_{ik} + \ddot{\alpha}_{ik}\right] (\alpha^{-1})^k_{\  j} (q^j + s^j)\,.
\ee
Taking the derivative $\partial/\partial q^j$, and comparing with \refeq{Mdot}, immediately yields
\be
\partial_{x,i}\partial_{x,j} \Phi = -\left[2\Hbg \dot{\alpha}_{ik} + \ddot{\alpha}_{ik}\right] (\alpha^{-1})^k_{\  j}\,.
\label{eq:Phialpha}
\ee
Thus, when restricting to non-relativistic matter, any given large-scale tidal perturbation $\Pi_{ij}(t)$ [\refeq{Pi}] can be treated as an effective anisotropic metric, with anisotropic scale factors determined by an ordinary differential equation (ODE). So far, the ``background'' scale factor $\abg(t)$ was merely a bookkeeping factor without physical relevance. We now identify it as the scale factor of the background cosmology with respect to which the tidal perturbation $\Pi_{ij}$ is defined. We then obtain
\be
\frac{d}{dt} \left( \abg^2 \dot\alpha_{ij} \right)
= - \frac32 \Omega_{m0} \Hbgn^2 \abg^{-1}(t) \alpha_i^{\  k} \Pi_{ki}(t)\,,
\label{eq:alphaEOM}
\ee
where we have rephrased the matter density $\rhobg \propto \abg^{-3}$ by using the Friedmann equation for $\abg$ and defining the density parameter $\Omega_{m0}$.

Now, we can use the freedom of rotating the simulation box with respect to the global coordinates, in such a way that $\alpha_{ij}$ becomes diagonal:
\ba
\nonumber
A_{ij}(t) =
\begin{pmatrix}
a_{1} & 0 & 0 \\
0 & a_{2} & 0 \\
0 & 0 & a_{3} \\
\end{pmatrix} &= \abg(t)  \, {\rm diag}\left( \alpha_1(t),  \alpha_{2}(t),  \alpha_3(t) \right)\,.
\ea
For simplicity we denote the diagonal elements of $\alpha_{ij}$ as $\alpha_{1,2,3}$. With this, \refeq{alphaEOM} becomes (see also \cite{2017arXiv171009881S})
\ba
\dot\alpha_i &= \abg^{-2} \eta_i \nonumber\\
\dot\eta_i &= - \frac32 \Omega_{m0} \Hbgn^2 \abg^{-1}(t) \alpha_i(t) \Pi_{ii}(t)\,,
\label{eq:alphaEOMi}
\ea
where $i \in \{ 1, 2, 3 \}$, and there is no summation over $i$.
\refeq{alphaEOMi} describes a set of ordinary differential equations that can be solved, for a given tidal field $\Pi_{ki}$, using standard methods. For this paper, we will always consider trace-free tidal perturbations $\Pi_{ij}\to K_{ij,L} \propto D(t)$ which follow linear evolution. We integrate the full nonlinear equations of motion \refeq{alphaEOMi},\footnote{We use the ODE solver included in the GNU scientific library (gsl) \url{https://www.gnu.org/software/gsl/}.\\
A standalone version of the algorithm to calculate the evolution of $\alpha_{i}, \eta_{i}$ for a tidal field can be found at \url{https://bitbucket.org/Avalon89/toolset/overview}} although this does not change the results significantly for the small amplitudes of $K_{ij,L}$ considered in this paper. Further, we parametrize the tidal tensor through (again, no summation over $i$ is implied)
\begin{align}
\Pi_{ii} = K_{ii,L} + \frac13 \delta_{L} &= D(t)\lambda_{i}\,,
\label{eq:tidal_kii}
\end{align}
where $\lambda_{i}$ is the amplitude today of the eigenvalues of the tidal tensor, and throughout we consider the case $\delta_{L}=0 \Leftrightarrow \lambda_1+\lambda_2+\lambda_3=0$.
If not noted otherwise we refer to $\lambda_{i}$ as the eigenvalue at $z = 0$.

\refeq{alphaEOM} simplifies further if we treat the tidal perturbation $\Pi_{ij}$ as a small parameter, decompose
\be
\alpha_{ij} = \d_{ij} + \hat\alpha_{ij}\,,
\ee
and work to linear order in $\Pi_{ij}, \hat\alpha_{ij}$. This leads to
\be
\frac{d}{dt} \left( \abg^2 \dot{\hat\alpha}_{ij} \right)
= - \frac32 \Omega_{m0} \Hbgn^2 \abg^{-1}(t) \Pi_{ki}(t)\,.
\label{eq:alphaEOMlin}
\ee
For reference, assuming a flat matter-dominated (Einstein-de Sitter) universe and adiabatic scalar perturbation such that $\Pi_{ki}(t) = \Pi_{ki}(t_0) \abg(t)$, one simply obtains
\be
\hat\alpha_{ij} \stackrel{\text{EdS, linear}}{=} -\Pi_{ij}(t)\,.
\label{eq:alphaEdS}
\ee
In the isotropic case $\hat\alpha_{ij} = \hat\alpha \d_{ij}$, we find that $\hat\alpha = -\d_{L}/3$, as follows from mass conservation at linear order in the standard, isotropic separate universe picture. We reiterate that our implementation and results are based on \refeqs{alphaEOM}{alphaEOMi}, which do not assume small tidal fields.

\begin{figure}
\includegraphics[width=0.47\textwidth]{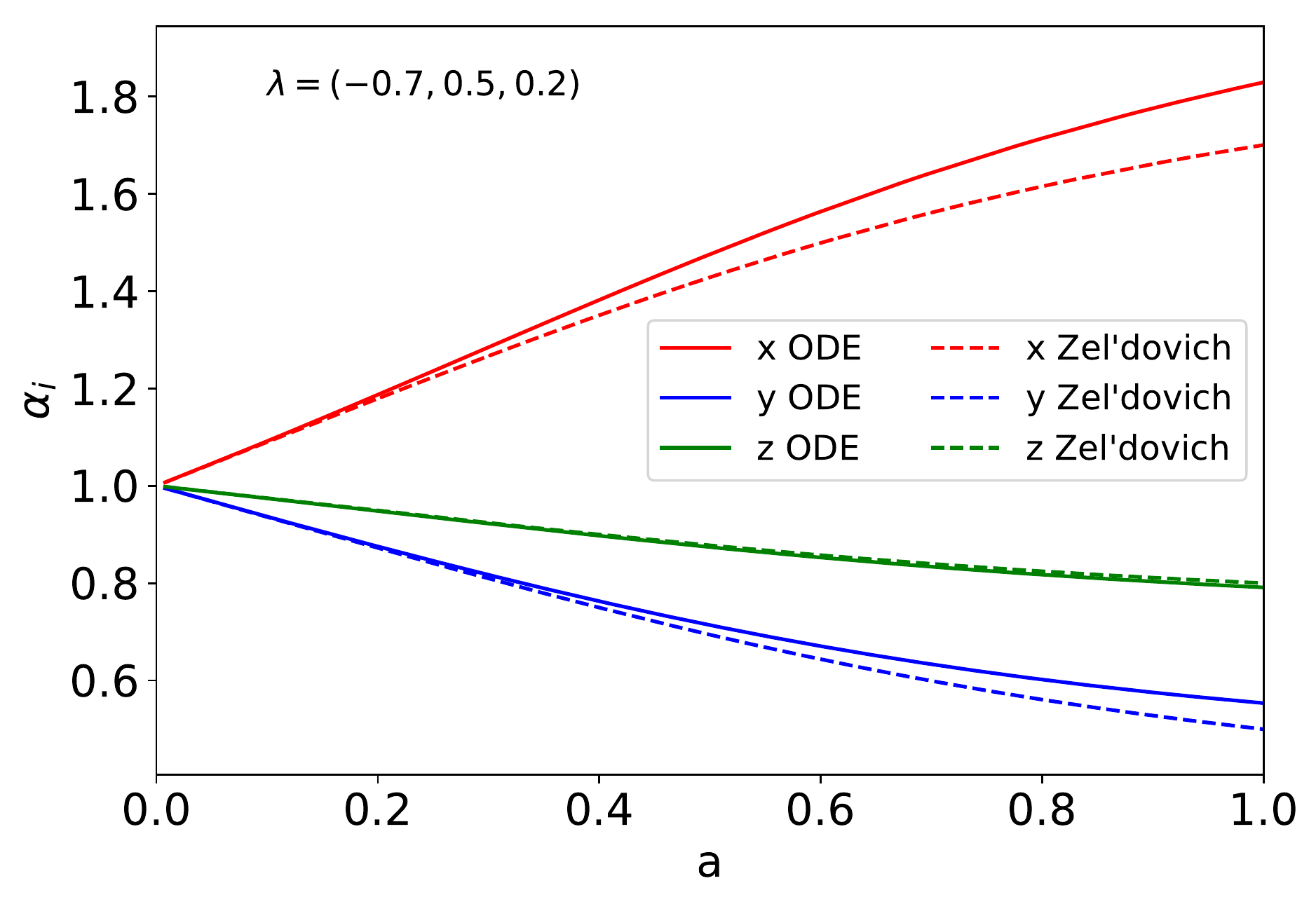}
\caption{The evolution of the relative scale factors $\alpha_i$ for an anisotropic region with deformation tensor eigenvalues $\lambda = (-0.7, 0.5, 0.2)$. The solid line represents the solution from the ordinary differential equation \refeq{alphaEOMi}, while the dashed lines represent the Zel'dovich approximation. The different axes are color coded according to the legend.}
\label{fig:evoaxes}
\end{figure}

In Fig. \ref{fig:evoaxes}, we show an example of the evolution of the three scale factors. Note that $\lambda_i$ are chosen to be quite large here for illustration. 
As expected, at early times the Zel'dovich approximation $\alpha_{i}(t) = 1 - \lambda_{i} D(t)$ (dashed lines) works well, while for later times the deviation from the numerical solution of the ODE's (solid lines) becomes significant. 
Fig. \ref{fig:evoaxes} shows that a negative $\lambda_{i}$ is stretching (increasing the expansion) while a positive $\lambda_{i}$ is squeezing the simulation box (reducing the expansion). 

Fig. \ref{fig:image_density_field} shows a visualization of the effect of a large-scale tidal field $K_{ij}$ on the structure in a small simulation box with 80 cMpc/h. We show the results both in comoving and in physical space.
In the comoving frame (upper panel), we see that the halos are stretched and squeezed forming ellipsoids while in the Eulerian frame (lower panel), where the box is rescaled according to the anisotropic scale factors $\alpha_{i}$, the halos appear spherical. This figure shows the result of the N-body implementation which we will describe below.
For the reminder of the paper we will drop the subscript $L$ and denote $K_{ij,L} \to K_{ij}$ and $\delta_{L} \to \delta$.

\section{Anisotropic N-body simulation}
\label{sec:Nbody}

After describing the model for the tidal field and how the anisotropic scale factor is evolved, we now discuss the modified equations of motion that are used to evolve the simulation, and how they are implemented in \gadget\, (Springel in prep.), an updated version of \Gadget 2 \citep{2005MNRAS.364.1105S}.

For this we define an anisotropic comoving frame which is related to physical coordinates by
\[
x_{i,\mathrm{phys}} = x_{i,\mathrm{com}} \cdot a_{i}\,.
\]
Note that our implementation is fully nonlinear in the tidal perturbation, although we will focus on an application to small (linear) tidal fields in this paper.

\subsection{Equation of Motion}

In the anisotropic case, the dynamics of the collisionless particles are described by the Hamiltonian\footnote{see \cite{2005MNRAS.364.1105S} for the isotropic case.}
\begin{equation}
H = \sum_{i}\left(\frac{1}{2 m_{i}}\sum_{k}\frac{p_{i,k}^2}{a_{k}(t)^2}\right) + \frac{1}{2}\sum_{i}\frac{m_{i}\phi(\vec{x}_{i})}{a(t)(\alpha_{x}\alpha_{y}\alpha_{z})}
\label{eq:Hamiltonian}
\end{equation}
with the canonical momentum $p_{i,k} = a_{k}^2m_{i}\dot{x}_{i,k}$, where the index $k \in [x,y,z]$ defines the axis. 
From the Hamiltonian \refeq{Hamiltonian} we obtain the equations for the change in momentum
\begin{equation}
\dot{\vec{p}}_{i} = - \frac{m_{i}}{2}\frac{\partial_{i} \phi}{a_{\mathrm{bg}}\alpha_{x}\alpha_{y}\alpha_{z}}\,,
\end{equation}
and the potential can be calculated by solving the Poisson equation, which in the anisotropic case is given as 
\begin{equation}
\underbrace{\sum_{i}\alpha_{i}^{-2} \partial_{i}^2}_{\equiv \nabla'^2} \phi = 4 \pi G \rho_{0,*}\delta \as{\abg^2}\,,
\label{eq:poisson_equation}	
\end{equation}
where the derivative $\nabla'^2 \phi$ is with respect to the new rescaled comoving coordinate, $\rho_{0,*} = \bar{\rho} / (a_x a_y a_z)$ is the mean density of the box, $\bar{\rho}$ is the mean density of the universe \as{without the imposed tidal field today} and $\delta$ describes the overdensity.
In Appendix \ref{sec:Appen} we give additional code-specific modifications that need to be done for the time step integrals used in the evolution operator in \gadget.

\subsection{Potential Calculations}
\label{sub:force}
The standard gravity solver used in \gadget is the TreePM algorithm, where the long range force is calculated using the PM method, and the short range is computed using the Tree.
We implemented a full TreePM algorithm to handle the anisotropic coordinates, but as our interest in this paper focuses on the large-scale structure and the weakly nonlinear regime, we use the code in a pure PM mode.
In a subsequent paper we will focus on small scale structure and halos using the full TreePM implementation.

On the particle mesh, the Poisson equation can be simply solved by Fourier transforming \refeq{poisson_equation}. That leads to 
\begin{align}
  \sum \alpha_i^{-2} k_i^2 \hat{\phi} &= 4 \pi G \rho_{0, *} \hat{\delta}\,, \\
  \hat{\phi} &= 4 \pi G \rho_{0,*} \frac{\hat{\delta}}{\sum \alpha_i^{-2} k_i^2}\,,\\
             &=: 4 \pi G \rho_{0,*} \hat{\delta} \hat{G}_*(\vec{k}) \label{eq:poisson_modified}
\end{align}
where $G_*$ denotes the Green's function. In practice the potential calculation on the particle mesh has to be modified by replacing the isotropic Green's function $\hat{G}(\vec{k}) = 1/(\sum k_i^2)$ by the anisotropic one $\hat{G}_*(\vec{k}) = 1/(\sum \alpha_i^{-2} k_i^2)$.

\section{Response}
\label{sec:theo}
Following the definition of the implementation, this section describes the property which we use to quantify the effect of the tidal field on the large-scale structure, which we will refer to as the tidal response function.
We follow the procedures of \cite{2017JCAP...06..053B} who defined a response function for the power spectrum, in particular the first order expansion set out in their Sec. 3.2.
The three dimensional power spectrum under the influence of a large-scale overdensity $\delta$ and an external tidal field $K_{ij}$ can be written as 
\begin{equation}
\label{eq:firstorderresponse}
P(\vec{k}) = P(k)\left(1 + R_{1}(k)\delta + R_{K}(k)\hat{k}_{i}\hat{k}_{j}K_{ij}\right)
\end{equation}
where $\hat{k}$ is a normalized $k$ vector such that $\left(\sum_{i} \hat{k}_{i}^2\right)^{1/2} = 1$ and $K_{ij}$ is the traceless tidal tensor. This expression is valid at linear order in $\delta$ and $K_{ij}$, and is independent of the wavelength of the large-scale perturbations, as long as it is much larger than $1/k$. Further, the response $R_{K}(k)$ is independent of the eigenvalues of the deformation tensor $K_{ij}$.

We write
\begin{equation}
\label{eq:response_tidal}
R_K(k) = G_K(k) - k \frac{P'(k)}{P(k)}\,,
\end{equation}
where $G_K(k)$ is the growth-only tidal response and $P(k)$ is the mean power spectrum, i.e. in the absence of any tidal effects. 
The growth-only tidal response is obtained when the modification of the power spectrum is measured in comoving coordinates.
\cite{2017PhRvD..95h3522A,2017JCAP...06..053B} derived at leading order in perturbation theory,
\begin{equation}
G_K^\text{LO}(k) = \frac87\,,
\label{eq:linear_approx_gk}
\end{equation}
which is valid on large-scales as $k\to 0$. We will compare our results to this result as a consistency test of the implementation. 
Otherwise higher order terms need to be taken into account and we would need to use at least the second order expansion to get an unbiased response function.
In the small amplitude regime for the tidal field the expansion above \refeq{firstorderresponse} is also valid on small scales where the structure is non-linear.

To avoid confusion, we refer to $R_{K}$ as the first order Eulerian response in the physical frame, while $G_{K}$ is the first order Lagrangian response in the comoving frame which in the linear regime is simply $G_{K} = 8/7$.
This Lagrangian response can be calculated by applying a standard power spectrum code keeping the box cubic (i.e., doing the power spectrum in the comoving frame).
The second term in \refeq{response_tidal} is a result of the coordinate transformation and can be evaluated full nonlinearly given a measurement (or fitting function) of the nonlinear isotropic matter power spectrum, and thus does not require anisotropic N-body simulations. 

Our simulations use a tidal tensor defined through the eigenvalues at $z=0$
\be
(\lambda_x, \lambda_y, \lambda_z) = \left(-\frac12, -\frac12, 1\right) \lambda_z \,.
\ee
We thus obtain
\begin{align}
\hat{k}^i\hat{k}^j K_{ij} &= D(t)\left(\lambda_{z}\hat{k}_{z}^2 - \frac{\lambda_{z}}{2}\hat{k}_{y}^2 - \frac{\lambda_{z}}{2}\hat{k}_{x}^2\right) \nonumber\\
&= \frac{D(t)\lambda_{z}}{2} \left(3\hat{k}_{z}^2-1\right) \nonumber \\
&= \lambda_{z} D(t) Y_{2}(\mu) 
\label{eq:tidal_special_case}
\end{align}
where $Y_2$ is the second-order Legendre polynomial and $\vec{\hat{k}}\cdot \vec{\hat{z}} = \mu = \hat{k}_{z}$ is the cosine of the angle between the k vector and the z axis.

\section{Response predictions}
\label{sec:halomodel}
We will consider two predictions for the response $G_K(k)$ on nonlinear scales. First, \cite{2017JCAP...06..053B} proposed that the shape of $G_K$ would follow that of the growth-only density response $G_1$ measured in \cite{2015JCAP...08..042W}, with a normalization chosen so that the correct low-$k$ asymptote is obtained:
\be
G_K(k) = \frac{12}{13} G_1(k)\,.
\ee
This was merely a simple ansatz to obtain numerical results for $R_K$ (and six further second-order response functions). 

Second, we derive the prediction for the nonlinear tidal response $G_K(k)$ in the halo model (see \cite{cooray/sheth} for a review), paralleling the derivation of the density response in \citet{takada/hu:2013}, \citet{posdeppk}, and \citet{2015JCAP...08..042W}. Adopting the notation of \cite{takada/hu:2013}, the halo model power spectrum, $P_{\rm HM}(k)$, is given by
\ba
P_{\rm HM}(k) =\:& P^{\rm 2h}(k) + P^{\rm 1h}(k) \label{eq:PkHM} \\
P^{\rm 2h}(k) =\:& \left[I^1_1(k)\right]^2 P_{lin}(k) \nonumber\\
P^{\rm 1h}(k) =\:& I^0_2(k,k)\,,\nonumber
\ea
where $P^{n {\rm h}}(k)$ denotes the $n$-halo term, 
\ba
I^n_m(k_1,\cdots k_m) \equiv \int& d\ln M\:n(\ln M) \left(\frac{M}{\rhob}\right)^m \, b_n(M)  \nonumber \\
&\times u(M|k_1) \cdots u(M|k_m)\,,
\label{eq:Inmdef}
\ea
and $n(\ln M)$ is the mass function (comoving number density per interval
in log mass), $M$ is the halo mass, $b_n(M)$ is the $n$-th order local bias 
parameter, $u(M|k)$ is the dimensionless Fourier transform of the halo 
density profile, for which we use the NFW profile \citep{NFW} and $P_{lin}$ is the linear power spectrum. We normalize
$u$ so that $u(M|k\to 0)=1$.  The notation given in \refeq{Inmdef}
assumes $b_0\equiv1$. $u(M|k)$ depends on $M$ through the scale radius
$r_s$, which in turn is given through the mass-concentration
relation. All functions of $M$ in \refeq{Inmdef}, along with $P_{lin}$, are also functions of
$z$ although we have not shown this for clarity. In the following,
we adopt the Sheth-Tormen mass function \citep{sheth/tormen} with
the corresponding peak-background split bias, and the mass-concentration
relation of \cite{bullock/etal}. The exact choice of the latter only has 
a small impact on the predictions which does not affect our conclusions.

Now consider the tidal response. First, the linear power spectrum changes according to
\be
P_{lin}(k) \to \left[1 + \frac87 \hat k^i \hat k^j K_{ij}\right] P_{lin}(k)\,.
\ee
Unlike the case of the response to a long-wavelength density perturbation, the halo number density is unchanged by a tidal field at linear order, since it is a scalar \citep{mcdonald/roy,MSZ,biasreview}. 
Thus, the only remaining effect to consider is a possible change in the halo profiles.

A good zeroth-order assumption is that the inner regions of halos are unaffected by the large-scale tidal field, since they virialize and decouple from large-scale perturbations at early times. Thus, the halo profiles are unchanged in \emph{physical} coordinates, which, in terms of our comoving coordinates, implies
\ba
u(M|\vk)\Big|_{K_{ij}} =\:& u\left(M \Big| \left[1 + K_{ij}\hat k^i \hat k^j\right] k \right)\,,
\ea
where we have expanded to linear order in $K_{ij}$ and used \refeq{alphaEdS}. Equivalently, since the NFW profile $u(M|k)$ is a function of $k r_s$, where $r_s = R_{\rm vir}(M)/c(M)$ is the scale radius and $c$ is the concentration, we can rephrase this rescaling in terms of the concentration:
\be
c(M)\Big|_{K_{ij}} = \left[1 + C_K K_{ij}\hat k^i \hat k^j\right] c(M)\,,
\ee
where we have introduced a constant $C_K$ to allow for a more general behavior. An unchanged halo profile in physical coordinates corresponds to $C_K=1$, since $c \propto 1/r_s$ is the inverse of a physical length. Clearly, we expect $C_K$ to be in the approximate range of $0 \lesssim C_K \lesssim 1$. 

Putting everything together, we obtain
\ba
G_K^\text{HM}(k) P_\text{HM}(k) =\:& \frac87 \left[I^1_1(k)\right]^2 P_{lin}(k)  \nonumber \\
 + C_K \Big[ 2 \left(I^1_1\right)_{,\ln c}&(k) I^1_1(k) P_L(k) + \left(I^0_2\right)_{,\ln c}(k,k) \Big]\,,
\ea
where
\ba
\left(I^1_1\right)_{,\ln c}(k) &= \int d\ln M\:n(\ln M) \left(\frac{M}{\rhob}\right) \, b_1(M) \left[\frac{\partial u(M|k)}{\partial\ln c}\right] \nonumber \\
\left(I^0_2\right)_{,\ln c}(k,k) &= 2 \int d\ln M\:n(\ln M) \left(\frac{M}{\rhob}\right)^2  u(M|k)  \nonumber \\
& \hspace*{1cm} \times \left[\frac{\partial u(M|k)}{\partial\ln c}\right]
\ea
are the derivatives of the relevant mass integrals with respect to the halo concentration. Note that both of these integrals scale as $k^2$ in the large-scale limit, so that the effect of the tidal field on halo profiles (in comoving units) is only relevant on small scales, as expected.

If the inner regions of halos indeed do not respond to the tidal field in physical space (corresponding to $C_K=1$), then we expect the Eulerian response to asymptote to zero at large $k$. Via \refeq{response_tidal}, this implies
\be
G_K(k) \stackrel{k\to\infty}{\longrightarrow} \frac{d\ln P(k)}{d\ln k}\,,
\ee
which is roughly $-2$. We will indeed see a change of sign in the simulation measurements of $G_K(k)$ on small scales.

\begin{figure}
\includegraphics[width = 0.47 \textwidth]{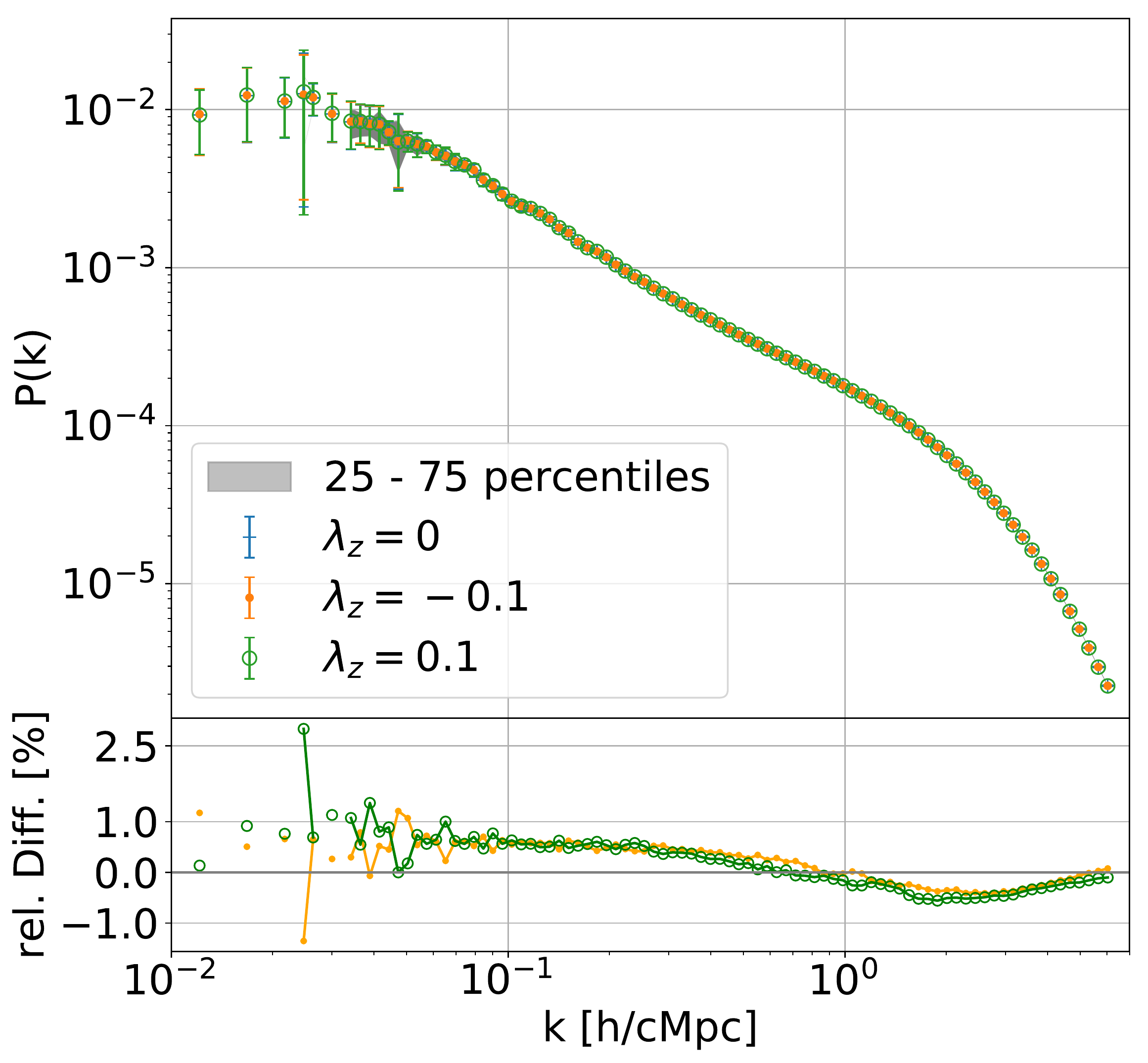}
\caption{The upper panel shows a comparison of the one-dimensional (i.e., angle-averaged) power spectrum for the three simulations using different tidal fields. The symbols show the mean with errorbar giving the rms scatter among the realizations.
Open circles show the case $\lambda = (-0.05, -0.05, 0.1)$, while black dots are for $\lambda = (0.05, 0.05, -0.1)$, and crosses are for runs with no tidal field.
The lower panel shows the relative difference between the simulation runs including a tidal field and the isotropic run without the influence from a tidal field.
The green line with open circles corresponds to the relative difference between the $\lambda_{z} = 0.1$ and the standard ($\lambda_{z} = 0$) while the orange line with dots represents the difference of the $\lambda_{z} = -0.1$ with the standard run.
We see that there is a small difference that is most likely from higher order terms $(K_{ij})^2$ that are expected of the order of $1$ percent.}
\label{fig:powerspectrum}
\end{figure}

\section{Simulation Setup}
\label{sec:sims}
This section is dedicated to the simulation setup and discusses the main characteristics of the runs.
To calculate the response of the power spectrum, we consider three choices for the imposed tidal field, two which differ in the sign of each eigenvalue, $\lambda_{i}$ and one for which all the $\lambda_{i}$ are zero.
For each, we consider 16 realizations of the initial density and velocity fluctuations at our starting redshift $z_{\rm init} = 127$.
These initial fluctuations are as expected in a fiducial flat $\Lambda$CDM cosmology with the Planck 2015 \citep{Planck2015} cosmological parameters, namely $\Omega_{m} = 0.308$, $\Omega_{\Lambda} = 0.692$, $\Omega_{b} = 0.04694$, $\sigma_{8} = 0.829$, and $h = 0.678$.
The evolution from each set of initial conditions is then calculated for each of our three choices of tidal field.
This allows us to get an estimate of the cosmic variance introduced by the finite size of our simulation volume.
\as{For convenience a}ll three simulations per response function estimate use exactly the same initial conditions, \as{which are computed using the standard Zel'dovich approach for an isotropic expanding universe without imposed tidal field. We thus neglect the influence of the large-scale tidal field at the starting redshift.}
This introduces small, \as{percent-level} artefacts which we address later in Sec. \ref{sec:results}.
\as{Further we will introduce a fully consistent model for the initial displacement in an ``separate universe'' under the influence of a tidal field in a subsequent paper presenting the modified TreePM.
This will eliminate the forementioned artefacts.}
The simulations are all run in pure PM mode with the modified Poisson equation \refeq{poisson_modified} and a grid for the PM of $2048^3$ cells.
Our simulation box has a size of 500 cMpc/h and we use $512^3$ particles.
With this, we get PM cells with a size of 244 ckpc/h for the PM mesh which sets our force resolution limit.

To separate effects coming from the tidal field from those due to large-scale density offsets, we limit ourself to simulations with $\sum_{i} \lambda_{i} = 0 = \delta_{L}$ (traceless), where the $\lambda_{i}$ are the eigenvalues of the linear deformation tensor and $\delta_{L}$ is the linear overdensity at $z = 0$ for the runs including a tidal field.
With this choice there are no contributions from $R_{1}$, since this part of the response is sourced by the overdensity.
Effects from a large-scale overdensity were already discussed in \citet{wagner/etal:2014} and \citet{2015JCAP...08..042W}.
Through the implementation of the tidal tensor presented here, we can simulate both a long-wavelength overdensity and the traceless tidal field.
Choosing all three eigenvalues $\lambda$ equal (isotropic) we have, at linear order, $\delta(z = 0) = \lambda_{1}+\lambda_{2}+\lambda_{3}$ and no tidal field.
Using this setup we ran an isotropic simulation to check the implementation and found good agreement with the growth-only response function $G_{1}$ from separate universe simulations.

\begin{figure}
\includegraphics[width = 0.47\textwidth]{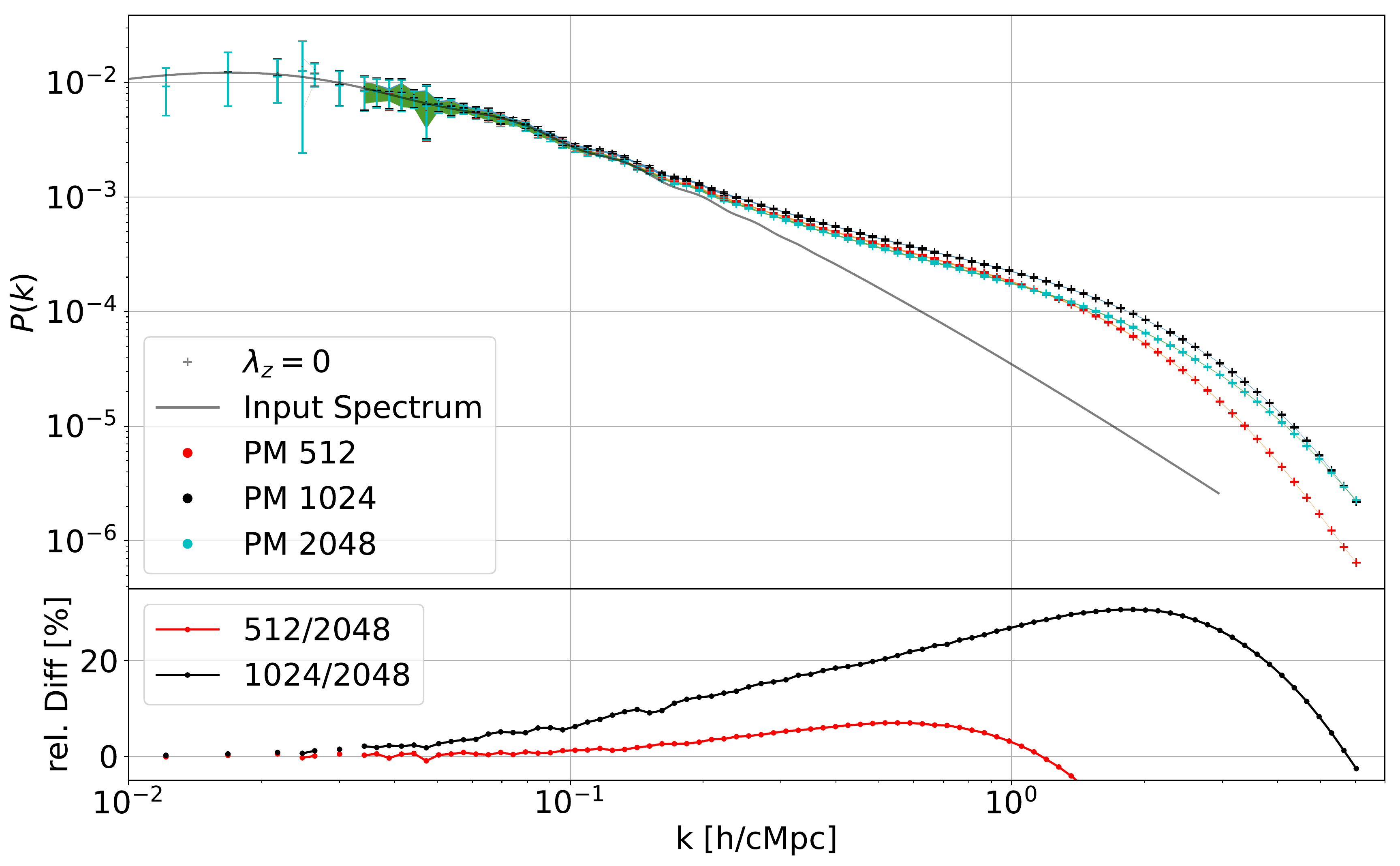}
\caption{Comparison of different PM grid resolutions on the convergence for the one-dimensional power spectrum. We only show the $\lambda = 0$, case as in the one-dimensional power spectrum all three $\lambda$ cases are identical. All runs are for a $512^3$ particle grid. 
In the upper plot we show the mean and rms scatter from the 16 realizations as symbols and errorbars respectively.
The fiducial PM resolution of 2048 is shown as cyan symbols, while the 512 and 1024 are shown as red and black. The grey line represents the input power spectrum for the initial conditions rescaled using the linear growth factor and is only plotted to $k \sim 3$.
In the intermediate regime ($10^{-1} \leq k [\mathrm{h/cMpc}] \leq 1$), the highest resolution drops in power below the smallest resolution.
The \as{intermediate} resolution is above both other resolutions for most of the scales and converges with the other resolutions at small $k$.
The lower panel shows the relative difference in percent between the smaller resolutions to the fiducial (2048) one.
One can see that, on intermediate scales, the case of 1024 cells per axis shows a significant difference to both the lowest and highest resolved runs.}
\label{fig:1D_power_convergence}
\end{figure}

The angle averaged (one-dimensional) power spectra for the three sets of simulations are shown in Fig. \ref{fig:powerspectrum}. The approximately one percent difference seen for simulations of either sign of $\lambda_z$ relative to the isotropic case is most likely due to higher-order terms in $K_{ij}$ (e.g. $(K_{ij})^2$), which are expected at the few-percent level. 

It is sufficient to consider the case where two eigenvalues are equal and the third is twice as large with the opposite sign so that the imposed tidal field is characterized by a single parameter.
For the numerical calculation of the response we symmetrically combine the three simulations run from each set of initial conditions (see below).
We thus end up with three simulations per measurement of $G_{K}$ and a total of 48 simulations for the full set.

\begin{figure}
\includegraphics[width=0.48\textwidth]{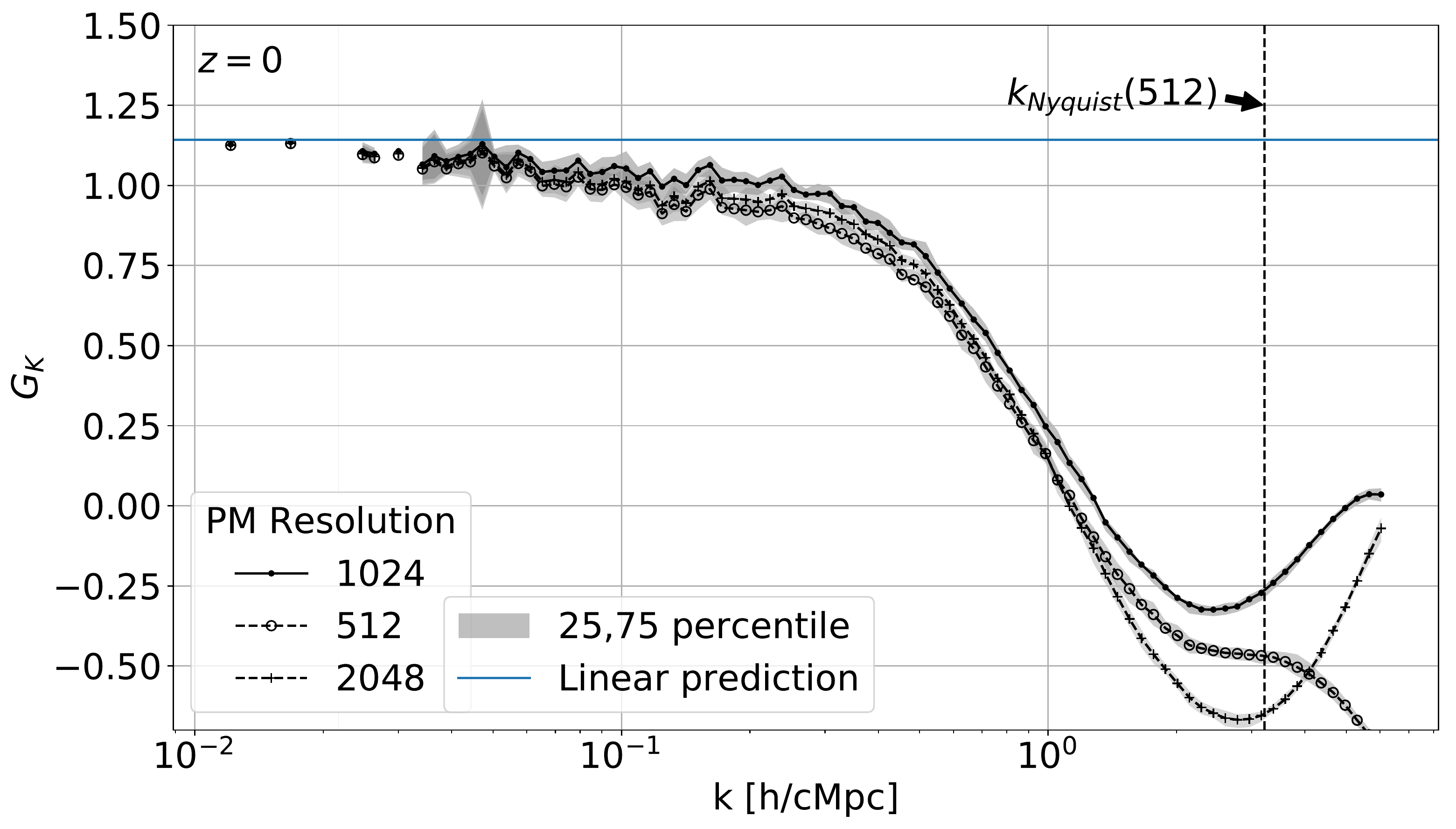}
\caption{Comparison of the mean for two sets of simulations with varying PM grid cells. \as{The filled and open circles represent the mean of the ($1024^3$) and small ($512^3$) grid respectively. Furthermore, the crosses show the fiducial $2048^3$ grid for the PM calculation}. The shaded areas represents the 25 to 75 percentiles for each set of 16 realizations.
The horizontal blue line in the upper panel shows the linear prediction from \citet{2017PhRvD..95h3522A}.
The vertical black dashed line ($k \sim 3.$) represents the particle Nyquist frequency which is also the Nyquist frequency for the smaller PM grid.}
\label{fig:comparison}
\end{figure}

To see the convergence and the scale where resolution effects become important, we run two more sets with the same particle number and initial conditions but with different PM resolution.
The first set has $512^3$ cells, while the second has $1024^3$ cells.
The comparison can be seen in Fig. \ref{fig:1D_power_convergence} and Fig. \ref{fig:comparison}, where we show the one-dimensional power spectrum and response function, respectively, for each force resolution.
The response function and power spectrum are computed using a Fourier grid of $1024^3$ cells that is unchanged for all PM resolutions.
We note that the power spectra and response functions of such pure PM simulations do not vary monotonically as the resolution is changed.
This produces the offset in the response function for the $1024^3$ cells run compared to the other two force resolutions.
This is also true for the one-dimensional power spectrum Fig. \ref{fig:1D_power_convergence}, which does show that changing the force resolution (PM grid size) influences the power spectrum.
The $512^3$ and $2048^3$ cell runs agree quite well on most scales while the $1024^3$ cell simulation is above both of them for nearly all scales.
We see that the response function agrees for all three force resolutions up to $k \sim 1.3$\, h/cMpc at roughly the ten percent level. However, larger departures are seen on smaller scales (higher $k$), most likely due to anisotropic force-softening effects.

\begin{figure*}
\includegraphics[width=0.9\textwidth]{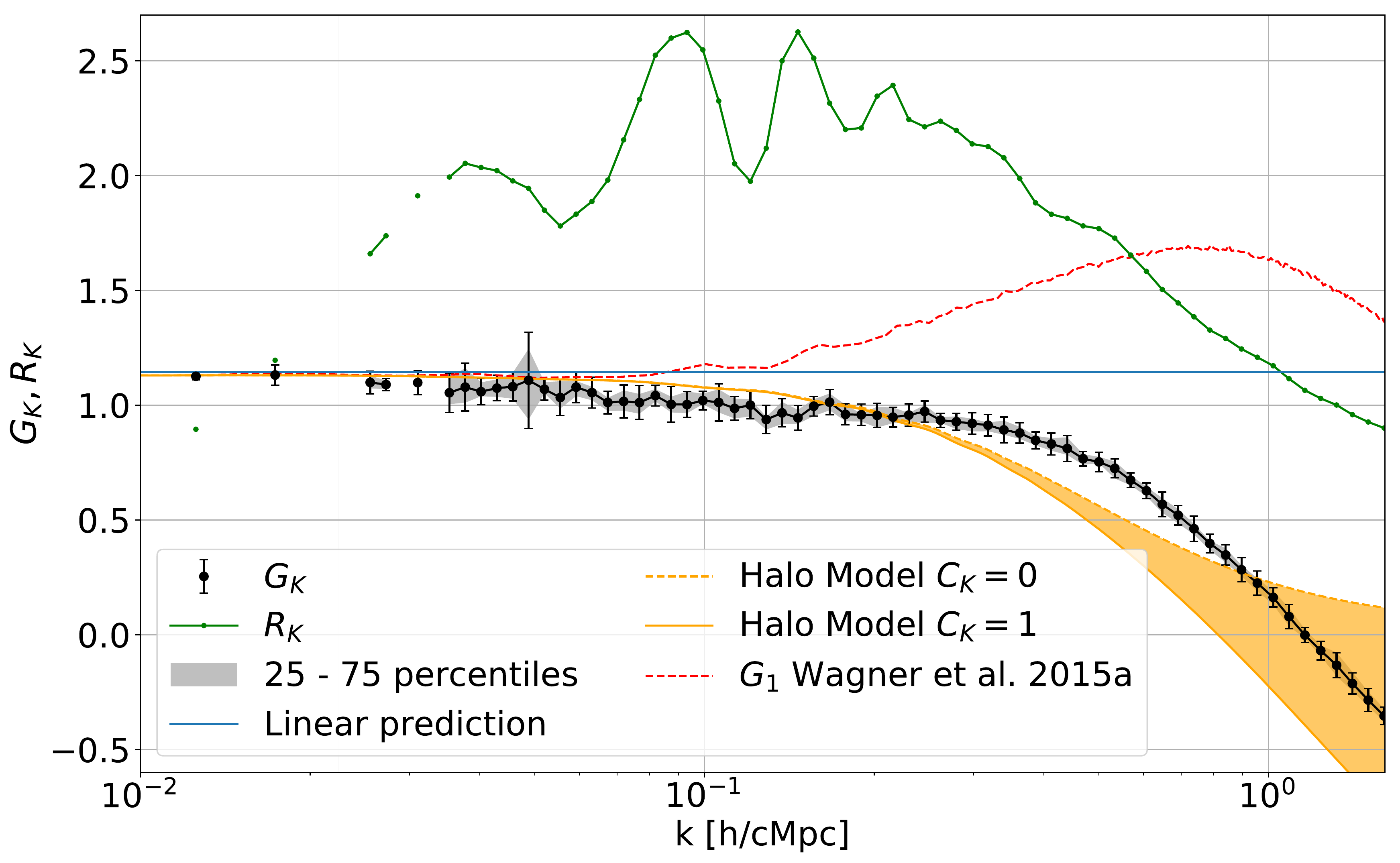}
\caption{The measured growth-only ($G_K$) and full ($R_K$) tidal response from our anisotropic N-body simulations. The black symbols show the $z = 0$ mean of the 16 realizations for $G_K$, and the errorbars represent the standard deviation\as{, where the larger errorbar around $k \sim 0.05 h/cMpc$ is due to the substantially fewer modes compared to neighboring bins}. The 25 to 75 percentiles are shown as grey band. The green line with dots shows $R_{K}$ constructed via \refeq{response_tidal} from the measured $G_{K}$ along with the logarithmic derivative of the power spectrum using CosmicEMU \citep{CosmicEmu}.  The horizontal solid blue line shows the perturbation-theory prediction from \citet{2017PhRvD..95h3522A}, while the red line represents the extrapolation from \citet{2017JCAP...06..053B}, using $G_{1}$ from \citet{2015JCAP...08..042W}. Finally, the halo model prediction described in Sec. \ref{sec:halomodel} is shown as orange shaded area, where the outlines are showing the result for $C_{K}=0$ and $C_K=1$.
}
\label{fig:HaloModelComp}
\end{figure*}

Given the unclear state of convergence of the PM results on very small scales
evidenced in Fig. \ref{fig:comparison}, we will limit ourselves to wavenumbers of $k \leq 2 \mathrm{h/cMpc}$ in our discussions and in our main results, Figs. \ref{fig:HaloModelComp} and \ref{fig:Response_z2}.

\section{Measurement}
\label{sec:Measurement}
We measure the response by computing the 3D power spectrum in the standard simulation without tidal field ($\lambda_{i} = 0\, \forall i$) and in two runs with symmetric eigenvalues $\lambda_{i,A} = -\lambda_{i,B}$ where the second index represents the simulation. Those three runs all originate from the same initial conditions. We use a Fourier grid for the power spectrum calculation of 1024 per axis which gives us a Nyquist frequency of $k_{Nyquist} = 6.43$\, h/cMpc.

To minimize effects from the initial conditions, we take the difference of simulations A and B and divide by the run without a tidal field.
Using this in \refeq{firstorderresponse} and taking into account the angular dependence by multiplying both sides with the second order Legendre polynomial $Y_{2}(\mu)$, we find
\begin{equation}
\label{eq:response_measure}
G_{K}(a) = \frac{\left\langle \left(P(\vec{k}|\lambda_{z,A}) - P(\vec{k}|\lambda_{z,B})\right) Y_{2}(\mu) \right\rangle}{\langle P(\vec{k}|\lambda_{z} = 0) Y^2_{2}(\mu) \cdot D(a) \underbrace{(\lambda_{z,A} - \lambda_{z,B})}_{\approx 2 \lambda_{z}} \rangle}
\end{equation}
where the $\langle ... \rangle$ denote angle-averaging, and $\lambda_{z,A} = -\lambda_{z,B} > 0$.
Here we weight both sides with $Y_{2}(\mu)$, as this optimally extracts the tidal response signal following \refeq{tidal_special_case}.
\as{The final step is to average the response \refeq{response_measure} over the 16 realisations.}
\refeq{response_measure} is unbiased up to corrections of order $(K_{ij})^2$, which, for our choice of $\lambda_z$, are on the order of 1 percent.

\section{Results}
\label{sec:results}
In this section, we will discuss our findings for the growth-only response from the N-body simulations.
From the measured response which is shown in Fig. \ref{fig:HaloModelComp}, we see the expected behavior on large scales, a constant following the perturbation-theory prediction of $G_{K} = 8/7$. On smaller scales we expect deviations from this due to nonlinear structure formation, which can be seen at $k \gtrsim 0.3\, \mathrm{h/cMpc}$ (scales of $2\pi/k \lesssim 21\, \mathrm{cMpc/h}$) for $z = 0$, where the response function starts to decline strongly.

We also compare our simulation results with the predictions for the extrapolated $G_{K}$ in \citet{2017JCAP...06..053B} which is given by $G_{K} = 12/13 G_{1}(k)$ and shown as red line in Fig. \ref{fig:HaloModelComp}. Unlike this extrapolation, the tidal response is not enhanced on intermediate scales but rather always suppressed with respect to the large-scale limit. We also show the predictions of the simple halo model (Sec. \ref{sec:halomodel}).
We see that the halo model does describe the features of the measured response. 
The predicted suppression in the non-linear regime is however somewhat steeper and occurs at somewhat smaller $k$ than seen in the N-body simulation. 
This is most likely a consequence of the simplistic assumptions made in the halo model, especially the assumption that all parts of the halo profile (the infall region as well as the inner core) respond equally weakly to the tidal field. Within the halo model, the key difference between $G_K$ and the density response $G_1$ is that a long-wavelength tidal field does not change the number of halos (at linear order), while a long-wavelength density perturbation does.

\begin{figure*}
\includegraphics[width=0.97\textwidth]{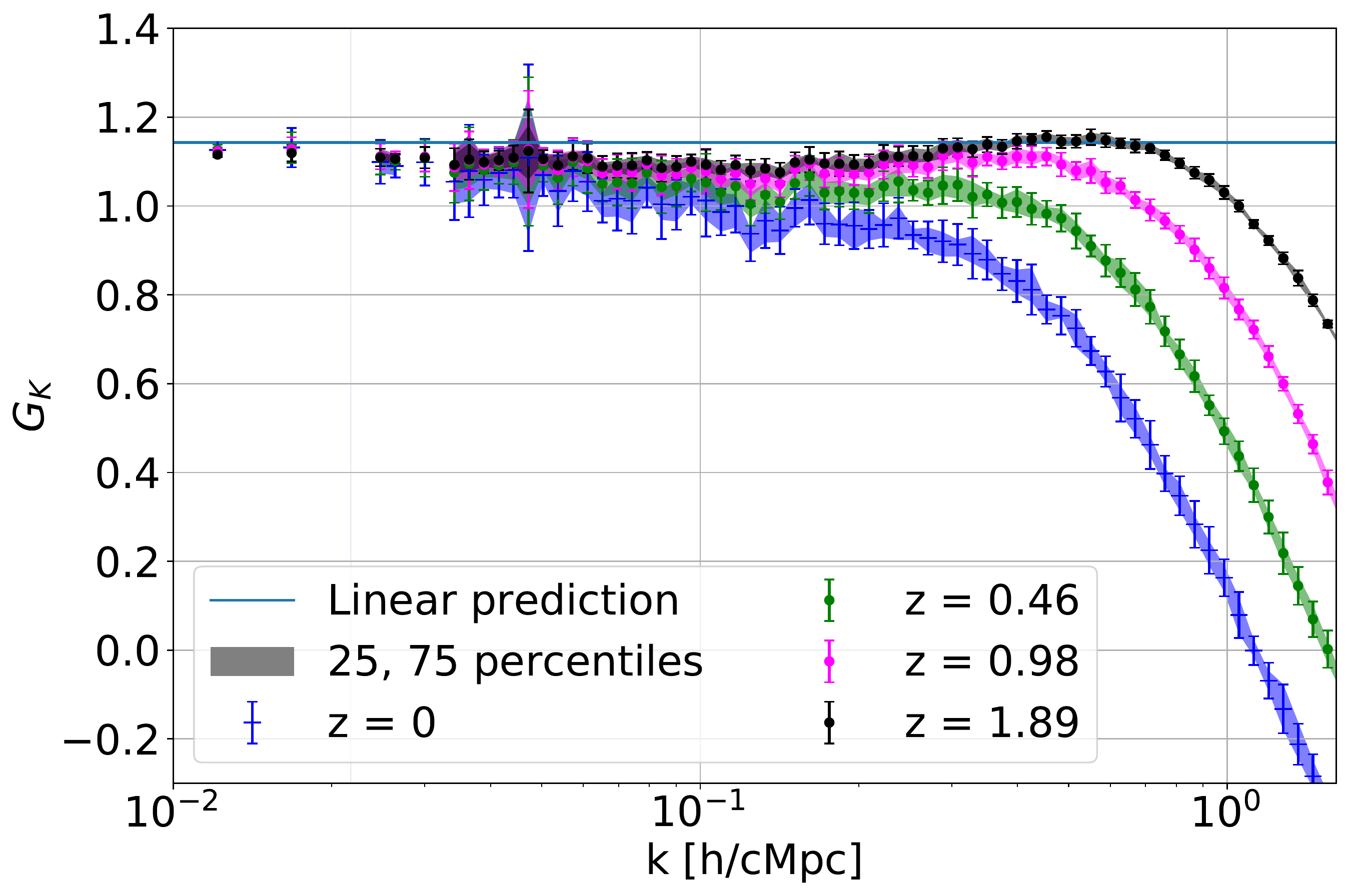}
\caption{Tidal response at higher redshifts: $z = 1.89$ (black symbols), $z = 0.98$ (magenta symbols), $z = 0.46$ (green symbols) and $z = 0$ (blue crosses). The blue line shows the perturbation-theory prediction from \citet{2017PhRvD..95h3522A}.
The shaded regions gives the 25 to 75 percentiles from the different realizations for the high redshift measurements.
We see that the non-linear regime evolves as expected, moving to lower $k$ values with decreasing redshift. 
There is a hint that at higher $z$ the response function at $k \sim 0.6$h/cMpc increases. However, this could be due to a decaying mode from the initial conditions.}
\label{fig:Response_z2}
\end{figure*}

The shift between the N-body simulation and the simple halo model suggests that the outer parts of halos as well as their environment, which are in the mildly non-linear regime, are significantly affected by a large-scale tidal field.
But as this regime is not determined by halos alone, this needs further investigation which will be the subject of further study focusing on intermediate and small scales as well as halo alignments.

On very small scales, $k \gtrsim 1.2$\, h/cMpc, Fig. \ref{fig:HaloModelComp} shows that the first order tidal response changes sign and becomes negative. Following the discussion in Sec. \ref{sec:halomodel}, this is expected if the inner regions of halos are only weakly affected by tidal fields. In the extreme limit, we expect $G_K(k) = d\ln P(k)/d\ln k \approx -2$. In order to study the behaviour of $G_K$ in this regime, the full TreePM code will be necessary.

Finally, using the first-order Lagrangian response function $G_{K}$ we can calculate the first-order Eulerian response function $R_{K}$ by subtracting the logarithmic derivative of the power spectrum derived from simulations without large-scale tidal field (see \refeq{response_tidal}). The tidal field response function in physical space $R_{K}$ is the one relevant for actual observations such as the power spectrum covariance. 
As the derivative of the power spectrum from simulations is normally noisy at large scales we choose to use emulated data for the one-dimensional power spectrum, calibrated on a much larger simulation data set and computed using the same cosmological parameters as our simulations. Specifically, we take the one-dimensional power spectrum from CosmicEMU \citep{CosmicEmu} and calculate its logarithmic derivative.
The result is shown in Fig. \ref{fig:HaloModelComp} as the green line.

In addition to the $z = 0$ response function discussed so far, we can look at earlier epochs.
We use four additional snapshots at $z \sim 0.5$, $z \sim 1$ and $z \sim 2$, shown in Fig. \ref{fig:Response_z2}. 
The wavenumber where the result deviates from the perturbation-theory prediction  shifts to higher k values with increasing redshift as expected (e.g. $k_{\mathrm{break}} \approx 0.7 \,\mathrm{h/cMpc}$ at $z \sim 2$). 
The deviation from the perturbation-theory prediction on the largest scales, which grows with redshift, is likely an artefact of the initial conditions that are taken to be identical for all simulations irrespective of the large-scale tidal field.
As a result, the initial response is forced to be $G_{K} = 0$ whereas it should, in fact be $G_{K} = 8/7$. This introduces a mismatch in the linear growing mode of order $D(z_{\rm init})/D(z)$, where $z_{\rm init}=127$ is the starting redshift of the simulations.

Further, at higher redshifts $z$ we see the appearance of a bump at scales of $k \sim 0.5 - 0.6$\, h/cMpc, which is most likely a decaying mode sourced by the unchanged velocity field in the initial conditions.
This is clearly a problem if one is interested in the early evolution of the response function, although the expected effect at $z = 0$ is at the percent level, as mentioned above. This issue will be addressed in a followup study.

\section{Summary}
\label{sec:summary}

We have simulated the effect of a large-scale tidal field on structure formation in a $\Lambda$CDM background cosmology using a modified version of \gadget with a PM force calculation adapted to an anisotropic background metric \refeq{anisoFRW}. The implementation is fully nonlinear in the amplitude of the anisotropy.

As a first applicaltion, we computed the first order growth-only tidal response function $G_{K}$ induced by this tidal field in the power spectrum, up to $k \simeq 2$ h/cMpc, using symmetric runs for the tidal field ($\lambda_{i,\mathrm{1}} = -\lambda_{i,\mathrm{2}}$), and recovered the predictions from perturbation theory  on large scales. 
Going to smaller scales, the extrapolation using the growth only response function $G_{1}$ for overdensities from \citet{2015JCAP...08..042W} does not fit our measurements.
In contrast to the interpolated solution, we find a suppression of the response on small scales compared to the large-scale value, which can be described approximately by the simple halo model in Sec. \ref{sec:halomodel}. The agreement with the simple halo model prediction is far from perfect however, signaling a tidal response of halos that is different in the inner and outer regions. A detailed analysis of this exceeds the scope of this paper, and will be the subject of an upcoming paper.
We also show the first order Eulerian response $R_{K}$ in Fig. \ref{fig:HaloModelComp} which was computed through the sum of $G_{K}$ and the logarithmic derivative of the isotropic power spectrum from CosmicEMU using the same background cosmology parameters as our simulations. This can now be used, for instance, in calculations of the covariance of the nonlinear matter and weak lensing shear power spectra.

\section{Acknowledgments}
The authors thank Volker Springel for help with \gadget and helpful discussion about the convergence, Alexandre Barreira for providing the data for $G_{1}$, and Titouan Lazeyras for data of the separate universe simulations and helpful discussions.
\as{AS is supported by DFG through SFB-Transregio TR33 ``The Dark Universe''.}
FS acknowledges support from the Starting Grant (ERC-2015-STG 678652) ``GrInflaGal'' from the European Research Council.

\bibliographystyle{mnras}
\bibliography{ref_paper}

\appendix
\section{Modified Evolution Operator}
\label{sec:Appen}
This appendix is meant to give some code specific details for the implementation of anisotropic scale factors.
In \gadget\,, the particles are evolved using a standard kick-drift-kick (KDK) leapfrog algorithm where the drift (D), kick (K) and the final evolution (E) operators are given as \citep{2005MNRAS.364.1105S}
\begin{align}
\label{eq:Evo_op}
E(\Delta t) &= K\left( \frac{\Delta t}{2} \right) D(\Delta t) K\left( \frac{\Delta t}{2} \right)\\
\label{eq:Drift_op}
D_{t}(\Delta t) &: 
\begin{cases}
\vec{p}_{n} &\rightarrow \vec{p}_{n} \\
\vec{x}_{n} &\rightarrow \vec{x}_{n} + \frac{\vec{p}_{n}}{m_{n}}\int_{t}^{t+\Delta t} \frac{dt}{a^2}
\end{cases}\\
\label{eq:Kick_op}
K_{t}(\Delta t) &:
\begin{cases}
\vec{x}_{n} &\rightarrow \vec{x}_{n}\\
\vec{p}_{n} &\rightarrow \vec{p}_{n} + \vec{f}_{n} \int_{t}^{t+\Delta t} \frac{dt}{a}
\end{cases}
\end{align}
where we have the drift and kick integrals and the force $\vec{f}_{n} = -\sum_{j} m_{n}m_{j}\nabla_{n}\phi(x_{nj})$.
In the two integrals contained in the operators we have factors of $a$ which need to be translated to the anisotropic case.
The transformed integrals to the actual time variable $a$ used in \gadget are
\begin{align}
I_{\mathrm{kick}} = I_1 = \int_{t}^{t+\Delta t} a^{-1} dt &= \int_{a(t)}^{a(t+\Delta t)} \frac{1}{a} \frac{1}{H(a) a} da \\
I_{\mathrm{drift}} = I_2 = \int_{t}^{t+\Delta t} a^{-2} dt &=  \int_{a(t)}^{a(t+\Delta t)} \frac{1}{a^2} \frac{1}{H(a) a} da 
\end{align}
which is the actual integral solved.
Now we can change the scale factors to the anisotropic scale factors and end up with the corresponding integrals, keeping in mind that the factor $H(a)a$ is a switch from time to scale factor in the integration and is therefore unchanged.
The new integrals in general depend now on the axis along which the integration is done, and we end up with six integrals (three for the kick and three for the drift) of the form
\begin{align}
I_{1} &= \alpha_{i}^{-1}\int_{a_{\mathrm{bg}}(t)}^{a_{\mathrm{bg}}(t+\Delta t)} \frac{1}{H(a_{\mathrm{bg}})a_{\mathrm{bg}}^2}\\
I_{2} &= \alpha_{i}^{-2} \int_{a_{\mathrm{bg}}(t)}^{a_{\mathrm{bg}}(t+\Delta t)} \frac{1}{H(a_{\mathrm{bg}})a_{\mathrm{bg}}^3}\,.
\end{align}
We moved the $\alpha_{i}$'s out of the integral under the assumption that the change in the time step is small, which is reasonable for most sensible cases of the tidal field\footnote{We also implemented a version where $I_{2}$ is integrated fully without that assumption showing only a very minor difference.}.
By moving $\alpha$ from the integral, we can absorb it for the kick integral in the force calculation and end up with the standard integral which reduces computation overhead.



\bsp
\label{lastpage}
\end{document}